\RequirePackage[2020-02-02]{latexrelease}
\documentclass[runningheads]{llncs}
\usepackage{import}
\usepackage{preamble}
\begin{document}
%
%
\title{Improvement Graph Convolution Collaborative Filtering with Weighted addition input}
%
%
\author{Tin T. Tran\inst{1,2}\orcidID{0000-0003-4252-6898} \and
V. Snasel\inst{1}\orcidID{0000-0002-9600-8319} }
\authorrunning{Tin T. Tran et al.}
\titlerunning{Improvement NGCF with Weighted addition input}
%
\institute{Faculty of Electrical Engineering and Computer Science, VŠB-Technical University of Ostrava, Ostrava-Poruba, Czech Republic
\email{\{trung.tin.tran.st,vaclav.snasel\}@vsb.cz} \and
Faculty of Information Technology, Ton Duc Thang University, Ho Chi Minh city, Vietnam\\
\email{trantrungtin@tdtu.edu.vn}}
\maketitle              

\begin{abstract}
Graph Neural Networks have been extensively applied in the field of machine learning to find features of graphs, and recommendation systems are no exception. The ratings of users on considered items can be represented by graphs which are input for many efficient models to find out the characteristics of the users and the items. From these insights, relevant items are recommended to users. However, user's decisions on the items have varying degrees of effects on different users, and this information should be learned so as not to be lost in the process of information mining.

In this publication, we propose to build an additional graph showing the recommended weight of an item to a target user to improve the accuracy of GNN models. Although the users' friendships were not recorded, their correlation was still evident through the commonalities in consumption behavior. We build a model WiGCN (Weighted input GCN) to describe and experiment on well-known datasets. Conclusions will be stated after comparing our results with state-of-the-art such as GCMC, NGCF and LightGCN. The source code is also included\footnote[1]{\url{https://github.com/trantin84/WiGCN}}.

\keywords{Recommender System  \and Collaborative signal \and Graph Neural Network \and Embedding Propagation.}
\end{abstract}
\section{Introduction}

Recommendation systems are an important research area of Information Systems. They are widely used in e-commerce, advertising and social media to give suggestions to the target user, and thus, improve the user experience. Based on previous interactions, such as purchases or reviews, of multiple users on items, the system will look for features and point out similar users and similar items to offer a set of recommendations. combine products and recommendations to target customers. Collaborative filtering (CF) is a method of assessing the similarity between users based on a series of past behavior that has been applied in \cite{5360986, 6779375} .In this method, the user and item features are represented by vectors. The distance of vectors in the calculating space represent how similar the users or items is. By this method, the items do not need to specify the attributes nor need to collect the user's personal information.

The user-item relationship matrix is usually a sparse matrix, because the number of items a user has purchased or is interested in accounts for only a very small portion of all items. Reducing the number of matrix dimensions helps to create embedded matrices of much smaller size but without losing the features of users and items \cite{10.1145/2556270}. Furthermore, the Numpy \cite{5725236} and TensorFlow libraries \cite{199317} can efficiently handle very large sparse matrices without taking up a lot of computer memory in execution.

In addition to traditional recommendations, social recommendations are systems that consider the social relation between users. Users can also be friends with each other in real life or on social networks, a friend's advice is always taken with higher trust. The methods outlined in \cite{1458205, SoReg} mined a relational database of friends to enrich recommendation information. However, a friendship database is not always available. Social networking services exploit user relationships only, while e-commerce sites have item review data and treat users as independent entities.

On the other hand, the user and item relations can also be naturally represented by graphs, and can be exploited by Graph neural network \cite{4700287}. The process of capturing collaborative signals can go through two stages, decomposer and combiner as suggested at \cite{wang2020multi}, where decomposer will capture signals from graph and then combiner will combine them into one unified embedding. High-order connectivity is found in the NGCF model \cite{NGCF} where propagating embedding is used when mining the user-item relationship graph. Recently, the LightGCN model has removed the weights when receiving the signal from the graph into the embedded vectors but still retains the accuracy of the algorithm \cite{he2020lightgcn}.

In this publication, we propose a model that uses Graph neural network to receive signals from two separate matrices, that is, the implicit matrix that records the user's attention on the item and the weight matrix. number of interactions between users. Although there is no data on social relationships between users, mutual interest in a set of items also constitutes strong signals between users. We also implement this model and perform experiments on common data sets to evaluate and compare with state-of-the-art algorithms.

\section{Related Works}
\subsection{Collaborative Filtering}
From the beginning, recommendation systems exploited the characteristics of all items as well as the preferences of users to find suitable items and make recommendations to them \cite{Jalili18}. For example, a movie can be categorized into action, comedy, romance, etc; a toy can have properties of color, shape, age, etc.\ With users, information such as age, gender, address of residence or education could also used to enrich the input data \cite{Pazzani99aframework}. However, with a very large number of items and frequent additions, it is difficult to find out and give items attributes, the systems need additive filtering methods. Collaborative filtering approach algorithms were used to find similarity between users or items without attribute information from them \cite{adomavicius2005incorporating}.

Collaborative filtering can be implemented by memory-based models \cite{6313742} or model-based models \cite{10.1145/138859.138867}. With memory-based model, the history of every user's rating on items will be recorded in a rating matrix. There are many ways to define a rating scale, which can be an integer value from 1 to 5, or an implicit rating. In a space made up of user set, item set, and rating values, each user is represented by a feature vector $e_u$, also as each item is represented by feature vector $e_i$. The algorithms will find out the distance between each pair of users (or pair of items for item-based algorithms) to find the neighbors and form the recommended results. The distance can be Cosine, Pearson, Euclidean similarity or Mean Squared Differences \cite{Candillier2007ComparingSC}. The weakness of these methods is that the input matrix is very large to be stored in the computer memory and the matrix is sparse because a user usually rates a few items.

In the model-based collaborative filtering systems, the algorithms will look for patterns in the learning data to create a model for future prediction \cite{doi:10.1137/1.9781611972726.4}. Matrix analysis techniques can be applied to reduce the dimensional of the user rating data matrix on items. Rating matrix can be decomposed into user feature matrix and item feature matrix with smaller size; but still ensure the accuracy of distance between users as well as between items, and more accuracy on recommendation prediction results \cite{5197422}, scalability and easier data learning process.

The evaluative measure will be selected appropriately for each algorithm \cite{10.1145/963770.963772}. A commonly evaluating method is to extract part of the data set to make a test set, the rest as a training set. The algorithm is applied on the training set to make predictions that are evaluated on the test set. The difference between the learning result and the actual data value shows the accuracy of the algorithm. This difference can be expressed in mean absolute error (MAE) or root mean square error (RMSE). Besides accuracy, coverage, scalability, learning time, memory consumption or interpretability are also an important criteria in evaluating the recommended system.

\begin{equation}
    MAE = (\frac{1}{n})\sum_{i=1}^{n}\left |  \widehat{y}_{i} - y_{i} \right |   \quad\text{and}\quad  RMSE = \sqrt{\frac{1}{n}\Sigma_{i=1}^{n}{\Big(\frac{\widehat{y}_i -y_i}{\sigma_i}\Big)^2}}
\end{equation}
where $\widehat{y}_i$ is predicted rating while $y_i$ is the value of that rating in the test set.

For implicit data, interactions between the user and the item are recorded binary, rather than rated as a specific value. Then, the algorithms will use the accuracy measure for the classification. The commonly used metrics are Precision and Recall \cite{Sarwar00, Sarwar00E}. Precision is the ratio between correct predictions on the test set, and Recall is the sensitivity of the algorithm, or the proportion of relational assertions that have been retrieved from the test set. The F1 score is used to balance accuracy and recall because the two are often opposite. Last but not least, discounted cumulative gain score (DCG) \cite{najork2008computing} assumes that judges have assigned labels to each result, and accumulates across the result vector a gain function G applied to the label of each result, scaled by a discount function D of the rank of the result, and it's normalized by the dividing DCG of an ideal result vector I.
\begin{equation}
    \begin{split}
        Precision &= \frac{TP}{TP+FP} \\
        Recall &= \frac{TP}{TP+FN} \\
        F1 &= \frac{2*Precision*Recall}{Precision+Recall}\\
        NDGC_u &= \frac{DCG_u}{CDG_{max}}
    \end{split}
\end{equation}
where $TP$ is true positive set of correctly predicting interaction exists between user and item. $FP$ is false positive set of missing prediction of interaction while $FN$ (False negative) shows that the predicted interaction doesn't exist on the test set.

\subsection{Graph Neural Networks}
Neural networks have had great success in setting up hidden layers to extract input data and identify features of the output data set \cite{gao2020deep}. A graph is a data structure that represents the relationship between entities, people, data, and attributes using a set of vertices $V$ to represent an object and an set of edges $E$ to represent the relationship. An edge of a graph can be undirected, representing only the relationship, or specifying the weight of the relationship, or might be a directed edge. Publication \cite{wu2021graph} have applied neural networks with graph input data and obtained positive results when vertex features are propagated and aggregated into each other during the learning process \cite{Wu2021}. Inheriting from Neural Network algorithms, GNN also uses multiple propagation layers and applies different aggregation and update solutions. The properties of the neighbor vertex are updated and the target vertex by pooling or attention mechanisms \cite{DBLP:journals/corr/abs-1903-07293}.

GCN is a method of collecting repeated information \cite{10.1145/3219819.3219890}. For example, when there is a similar item of interest to many of the target user's friends, it should be recommended with a higher degree than other items. GraphSAGE \cite{DBLP:journals/corr/HamiltonYL17} embedded inductive for each vertex of the graph and learned the topological structure of the graph as well as the effect of vertices on neighboring vertices. This method not only focus on feature-rich graphs but also make use of structural features of all vertices.

Multi Component Graph Convolutional \cite{wang2020multi} viewed a two-component input graph by separating the vertices representing the user as one group and the vertices representing the item as the other group. The features of the users are calculated to embed each other and so are the features of the items. The learning process is divided into two main stages: decomposer and combiner. In which, the decomposer component will obtain some $M$ latent flows from the user-item relation matrix through $M$ transformation matrices and aggregate into the user (or item) vertices. The combiner component will combine the feature vectors of the items on a particular user and those of the users on a particular item, and finally the system obtains a predictive rating on the output.

Assuming that the influence of the users is different depending on the distance between the vertices in the graph, Neural Graph Collaborative Filtering \cite{NGCF} generates hop-by-hop propagation classes in the input graph. The $k^{th}$ layer of propagation, called $k-order$ propagation, receives messages from items to a user as well as attention signals from that user's neighbors. The number of propagation layers $k$ can be considered as the input parameter and the value $k=3$ is considered to be the most optimal. The loss function is built from the difference of the predicted and actual evaluation values of the test set. The learning process to minimize the loss function takes place after a number of iterations depending on the learning rate, the size of the embedding matrices as well as the size of the learning data. In a subsequent version of NGCF, LightGCN \cite{he2020lightgcn} removed the characteristic transition weight matrices during propagation, and a non-linear activation function to remove negative effects on objects in the NGCF propagation process.

\section{Proposed method}
In this chapter, we will present a recommendation model to learn the user and item characteristics at the input, and then predict the recommended outcome at the output. We also suggest the metric for the recommended results and how to build functions in the hidden layers as well as the loss function in the learning process.

\subsection{Adjacent graph and Weighted references matrix}
    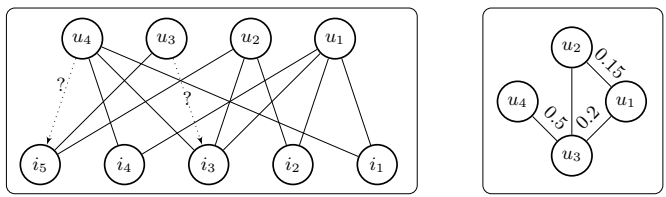
\begin{figure}[ht] 
        \centering
        \scalebox{0.8}{
       \begin{tikzpicture}
\tikzstyle{main} = [draw, circle, thick]
\tikzstyle{block} = [rectangle, draw, fill=blue!20,     text width=5em, text centered, rounded corners,     minimum height=1em]
\tikzstyle{line} = [draw, -latex']
\tikzstyle{box} = [rectangle, draw, width=1em,
                        height = 0.4em]
\tikzstyle{list} = [rectangle split, rectangle split parts=6,
                 rectangle split vertical, draw,
                 align=center,
                 text width=7mm, 
                 minimum height=9mm, 
                 inner sep=1mm, on chain]

\tikzset{
ui/.pic={
\node[main] (1) at (8, 0) {$i_1$}; 
\node[main] (2) at (7, 3) {$u_1$}; 
\node[main] (3) at (6, 0) {$i_2$}; 
\node[main] (4) at (5, 3) {$u_2$}; 
\node[main] (5) at (4, 0) {$i_3$}; 
\node[main] (6) at (3, 3) {$u_3$}; 
\node[main] (7) at (2, 0) {$i_4$};
\node[main] (8) at (1, 3) {$u_4$};
\node[main] (9) at (0, 0) {$i_5$};
\draw (1) -- (2); 
\draw (1) -- (8);
\draw (3) -- (2); 
\draw (3) -- (4);
\draw (5) -- (2); 
\draw (5) -- (4);
\draw (5) -- (8);
\draw (7) -- (2);
\draw (7) -- (8);
\draw (9) -- (4); 
\draw (9) -- (6);
\path [line, dotted] (8) -- node [text width=5.5cm,midway,above=0.2em,align=center ] {?} (9);
\path [line, dotted] (6) -- node [text width=5.5cm,midway,above=-0.5em,align=center ] {?} (5);
}
} 
\tikzset{
uu/.pic={
\node[main] (11) at (2, 1) {$u_1$}; 
\node[main] (12) at (1, 2) {$u_2$}; 
\node[main] (13) at (1, 0) {$u_3$}; 
\node[main] (14) at (0, 1) {$u_4$};
\draw [postaction={decorate},
      decoration={raise = 0.5 em, text along path,
                  text={0.15},
                  text align={center}
                  }](12) -- (11); 
\draw [postaction={decorate},
      decoration={raise = 0.5 em, text along path,
                  text={0.2},
                  text align={center}
                  }](13) -- (11);
\draw [postaction={decorate},
      decoration={raise = 0.3 em, text along path,
                  text={},
                  text align={center}
                  }](13) -- (12); 
\draw [postaction={decorate},
      decoration={raise = 0.5 em, text along path,
                  text={0.5},
                  text align={center}
                  }](14) -- (13);}
} 
\tikzset{
sq/.pic={
\node[list] (b1) [minimum height=10em, minimum width=30 em] at (0,0) {};
\node[box] (b2) [below=-\pgflinewidth of b1] {};
\node[box] (b3) [below=-\pgflinewidth of b2] {...};
\node[box] (b4) [below=-\pgflinewidth of b3] {};
\node[box] (b5) [below=-\pgflinewidth of b4] {};
}
}

\node (UI) [block, fill=none,minimum height=9.5em, minimum width=21 em ] at (0,0) {};

\node (UU) [block, fill=none,minimum height=9.5em, minimum width=9.2em] at (6,0) {};

\pic (ui)[scale=0.7,below=of UI.west, xshift=2.5em, yshift=1.25em] {ui};
\pic (uu)[scale=0.9,below=of UU.west, xshift=2em, yshift=1.5em] {uu};

\end{tikzpicture} 
        }
        \caption{Bi-parties users-items graph and Graph of users' influences.}
        \label{fig:my_label}
    \end{figure}
Due to history of interaction between users and items, we can model a bi-parties graph $G=\{V, E\}$, where set of vertices $V=\{U, I\}$ is union from set of users and set of items; the E contains edge $e_{i,j}=(u_i, i_j)$ if user u had interaction on item i. From the bi-parties graph G, we can define a matrix $R  \subseteq U  \times  I$ and it's elements has a value of
    \begin{equation}
         R_{i,j}=\begin{cases}
            1 & \text{user $u_i$ has an interaction on item $i_j$}\\
            0 & \text{otherwise}
         \end{cases} 
    \end{equation}
The weighted user references matrix $W_U = R \times R^T$ shows how many common items that user $u_i$ and $u_j$ have interaction by value of $ W_{U_{i,j}} $. Because the number of interactive items of each user are different, matrix $W$ should by normalized by least absolute deviations. Similarly, the weighted item references matrix $W_I = R^T \times R$ indicates how many users the same two items were referenced by.

According to \cite{NGCF}, the Laplacian matrices for user-item relation should be formed in propagation rule because all embedding vectors could be updated simultaneously. The first input $\Gamma$ is a square matrix whose dimensions each are the sum of the number of users and the number of items. 
\begin{equation}
     \Gamma =  D^{ -\frac{1}{2}}AD^{ -\frac{1}{2}} \quad\textrm{and}\quad  A =  \begin{bmatrix}0 & R \\ R^ \top   & 0 \end{bmatrix} \label{Fgamma}
\end{equation}
where $0$ is all-zero matrix and $D$ is the diagonal degree matrix, the $i^{th}$ element $D_{i,i}=|N_i|$. The second input is matrix $\Delta$, which has the same size with $\Gamma$, collects Weighted references values of both users and items.
\begin{equation} \label{Fdelta}
     \Delta =  D^{ -\frac{1}{2}}BD^{ -\frac{1}{2}} \quad\textrm{and}\quad B =  \begin{bmatrix}W_{U} & 0 \\ 0    & W_{I} \end{bmatrix}
\end{equation}

\subsection{Embedding layers}
In publication \cite{he2017neural}, authors used two vector for user's latent and item's latent, the size of the vectors for each user and each item can be taken as an input parameter. Thus, we define users embeddings and items embeddings as $e_u  \in \R^d$ and $e_i \in \R^d$ with $d$ is size of embeddings vector. In the first loop of propagation, embedding layer $E^{(0)}$ need an initial state of He normal weight initialization method which has good performance with Rectified Linear Unit activation function \cite{DBLP:journals/corr/abs-2004-06632}.
\begin{equation}
    \begin{split}
            E_{user}&=[e_{u_1},\ldots,e_{u_n}] \\
    E_{item}&=[e_{i_1},\ldots,e_{i_m}] \\
    E &= [E_{user} | E_{item}]
    \end{split}
\end{equation}

With input matrices in \eqref{Fgamma} and \eqref{Fdelta}, the embedding should be trained though each round of propagation as
\begin{equation}
    E^{(k)} = LeakyReLU \Big[ \Gamma E^{(k-1)}W_1^{(k-1)}  + \Delta \Gamma E^{(k-1)}W_2^{(k-1)}                + b^ {(k-1)} \Big]
\end{equation}
where $W_1^{(k-1)}$, $W_2^{(k-1)}$ are the trainable transformation matrices and $b^{k-1}$ is bias vector.

\subsection{Propagation process}
Collaborative filtering method will capture signals inside the graphs' structure and training the embeddings of both users and items. As message-passing model of GNN \cite{pmlr-v80-xu18c,NIPS2017_5dd9db5e}, we extract the signals and then make an aggregation for the embedding at output.
\subsubsection{Extract signals} With a user-item pair, the the signals sent out by an users and an item into target user is respectively
\begin{equation}
    m_{u \leftarrow u} = W_1e_u      
\end{equation}
\begin{equation}
    m_{u \leftarrow i}= \frac{1}{ \sqrt{| N_u ||N_i| } } *\Big( W_1 e_i  +\Delta W_2 e_i \Big)
\end{equation}

where $W_1$ and $W_2$ are trainable weight matrices. $\Delta$ is the Weighted addition matrix at the input. $|N_i|$ and $|N_u|$ are is the force of the user and item set, respectively.

\subsubsection{Aggregate signal}
With messages received from neighborhood of a user $u$, we aggregate all of them to refine representation of $u$.
\begin{equation}
    e^{(1)}_u = LeakyReLU \big(m_{u \leftarrow u}+ \sum_{i \in  \aleph _u} m_{u \leftarrow i}  \big) 
\end{equation} \label{ct1}
where $e^{(1)}_u$ present embedded latent of user $u$ after the first propagation layer. LeakyReLU is the activation function allows messages to encode both positive and a little negative signals \cite{Maas2013RectifierNI}. The relation of users is taken into consideration by two sources that are $ m_{u \leftarrow u}$ and matrix $\Delta$.

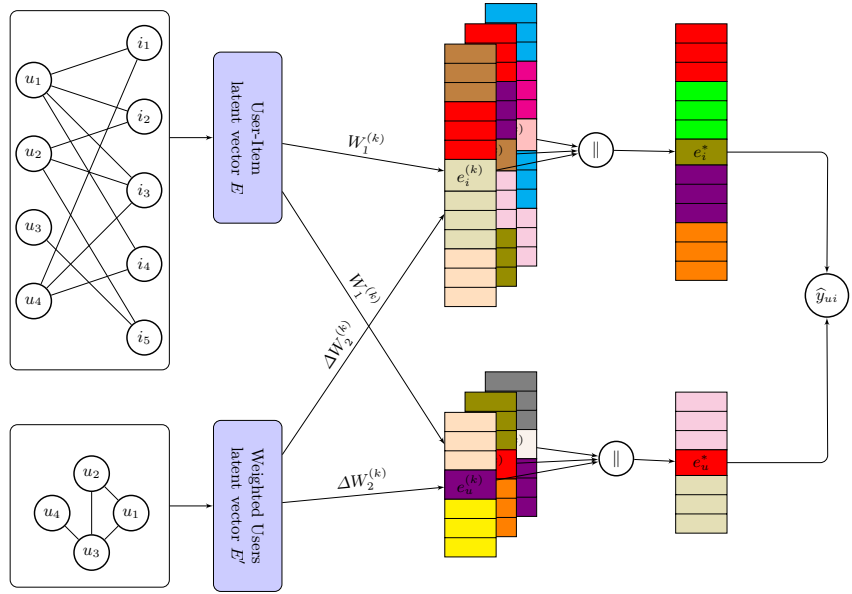
\begin{figure}[ht] 
\centering
\scalebox{.7}{
\begin{tikzpicture}
\tikzstyle{main} = [draw, circle, thick]
\tikzstyle{block} = [rectangle, draw, fill=blue!20,     text width=8em, text centered, rounded corners,     minimum height=4em]
\tikzstyle{line} = [draw, -latex']
\tikzstyle{box} = [rectangle,minimum width=3em,minimum height=1em, draw, inner sep=0.1cm]
\tikzstyle{boxs} = [rectangle split, rectangle split parts=3,minimum width=3em,minimum height=1em, draw, inner sep=0.1cm]
\pgfmathdeclarerandomlist{MyRandomColors}{%
    {red}%
    {red!25}%
    {magenta}%
    {magenta!25}%
    {olive}%
    {olive!25}%
    {brown}%
    {brown!10}%
    {violet}%
    {violet!25}%
    {gray}%
    {purple}%
    {yellow}%
    {orange}%
    {orange!25}%
    {cyan}%
    {green}%
}%

\tikzset{
ui/.pic={
\node[main] (1) at (3, 8) {$i_1$}; 
\node[main] (2) at (0, 7) {$u_1$}; 
\node[main] (3) at (3, 6) {$i_2$}; 
\node[main] (4) at (0, 5) {$u_2$}; 
\node[main] (5) at (3, 4) {$i_3$}; 
\node[main] (6) at (0, 3) {$u_3$}; 
\node[main] (7) at (3, 2) {$i_4$};
\node[main] (8) at (0, 1) {$u_4$};
\node[main] (9) at (3, 0) {$i_5$};
\draw (1) -- (2); 
\draw (1) -- (8);
\draw (3) -- (2); 
\draw (3) -- (4);
\draw (5) -- (2); 
\draw (5) -- (4);
\draw (5) -- (8);
\draw (7) -- (2);
\draw (7) -- (8);
\draw (9) -- (4); 
\draw (9) -- (6);}
} 
\tikzset{
uu/.pic={
\node[main] (11) at (2, 1) {$u_1$}; 
\node[main] (12) at (1, 2) {$u_2$}; 
\node[main] (13) at (1, 0) {$u_3$}; 
\node[main] (14) at (0, 1) {$u_4$};
\draw (11) -- (12); 
\draw (11) -- (13);
\draw (13) -- (12); 
\draw (13) -- (14);}
} 
\tikzset{
sq/.pic={
\pgfmathrandomitem{\RandomColor}{MyRandomColors} 
\node[boxs, fill=\RandomColor] (b1) at (0,0) {};
\pgfmathrandomitem{\RandomColor}{MyRandomColors} 
\node[boxs, fill=\RandomColor] (b2) [below=-\pgflinewidth of b1] {};
\pgfmathrandomitem{\RandomColor}{MyRandomColors} 
\node[box, fill=\RandomColor] (b3) [below=-\pgflinewidth of b2] {$e_i^{(k)}$};
\pgfmathrandomitem{\RandomColor}{MyRandomColors} 
\node[boxs, fill=\RandomColor] (b4) [below=-\pgflinewidth of b3] {};
\pgfmathrandomitem{\RandomColor}{MyRandomColors} 
\node[boxs, fill=\RandomColor] (b5) [below=-\pgflinewidth of b4] {};
}
}

\tikzset{
squ/.pic={
\pgfmathrandomitem{\RandomColor}{MyRandomColors} 
\node[boxs, fill=\RandomColor] (b1) at (0,0) {};
\pgfmathrandomitem{\RandomColor}{MyRandomColors} 
\node[box, fill=\RandomColor] (b2) [below=-\pgflinewidth of b1] {$e_u^{(k)}$};
\pgfmathrandomitem{\RandomColor}{MyRandomColors} 
\node[boxs, fill=\RandomColor] (b3) [below=-\pgflinewidth of b2] {};
}
}

\tikzset{
sqx/.pic={
\pgfmathrandomitem{\RandomColor}{MyRandomColors} 
\node[boxs, fill=\RandomColor] (b1) at (0,0) {};
\pgfmathrandomitem{\RandomColor}{MyRandomColors} 
\node[boxs, fill=\RandomColor] (b2) [below=-\pgflinewidth of b1] {};
\pgfmathrandomitem{\RandomColor}{MyRandomColors} 
\node[box, fill=\RandomColor] (b3) [below=-\pgflinewidth of b2] {$e_i^*$};
\pgfmathrandomitem{\RandomColor}{MyRandomColors} 
\node[boxs, fill=\RandomColor] (b4) [below=-\pgflinewidth of b3] {};
\pgfmathrandomitem{\RandomColor}{MyRandomColors} 
\node[boxs, fill=\RandomColor] (b5) [below=-\pgflinewidth of b4] {};
}
}

\tikzset{
squx/.pic={
\pgfmathrandomitem{\RandomColor}{MyRandomColors} 
\node[boxs, fill=\RandomColor] (b1) at (0,0) {};
\pgfmathrandomitem{\RandomColor}{MyRandomColors} 
\node[box, fill=\RandomColor] (b2) [below=-\pgflinewidth of b1] {$e_u^*$};
\pgfmathrandomitem{\RandomColor}{MyRandomColors} 
\node[boxs, fill=\RandomColor] (b3) [below=-\pgflinewidth of b2] {};
}
}

\node (UI) [block, fill=none,minimum height=21em, minimum width=9.3em ] at (0,6) {};
\node (UU) [block, fill=none,minimum height=9.5em, minimum width=9.3em] at (0,0) {};
\pic (ui)[scale=0.7,below=of UI.west, xshift=2em, yshift=-6.4em] {ui};
\pic (uu)[scale=0.75,below=of UU.west, xshift=3.3em, yshift=1.9em] {uu};
\node[block, rotate=-90, minimum width=10em] at (3,7) (UIGNN) {User-Item latent vector $E$};
\node[block, rotate=-90, minimum width=10em] at (3,0) (UUGNN) {Weighted Users latent vector $E'$};

\path [line] (UU) -- (UUGNN);
\path [line] (UI.east) ++(0,1)-- (UIGNN);
\pic [local bounding box=A11][scale=0.7] at (8,2,0) {squ};
\pic [local bounding box=A12][scale=0.7] at (8,2,1) {squ};
\pic [local bounding box=A13][scale=0.7] at (8,2,2) {squ};

\path [line, postaction={decorate},
      decoration={raise = 0.5 em, text along path,
                  text={${\Delta W^{(k)}_2}$},
                  text align={center}
                  }] (UUGNN) -- (A13) ;
\path [line, postaction={decorate},
      decoration={raise = 0.5 em, text along path, 
                  text={${W^{(k)}_1}$    \space\space\space \space\space\space\space},
                  text align={center},
                  }]  (UIGNN) -- (A13);

\pic [local bounding box=B11,scale=0.7] at (8,9,0) {sq} ;
\pic [local bounding box=B12,scale=0.7] at (8,9,1) {sq} ;
\pic [local bounding box=B13,scale=0.7] at (8,9,2) {sq} ;
\path [line, postaction={decorate},
      decoration={raise = 0.5 em, text along path,
                  text={${\Delta W^{(k)}_2}$ \space\space\space \space\space\space\space},
                  text align={center}
                  }]  (UUGNN) -- (B13) ;
\path [line, postaction={decorate},
      decoration={raise = 0.5 em, text along path,
                  text={${W^{(k)}_1}$},
                  text align={center}
                  }]  (UIGNN) -- (B13);

\pic [local bounding box=Af][scale=0.7] at (12,2,1) {squx};
\pic [local bounding box=Bf,scale=0.7] at (12,9,1) {sqx} ;

\node [main] at (10, 7.15,1) (conB) {$ \parallel $};
\node [main] at (10, 0.9,0) (conA) {$  \parallel $};

\path [line] (A11) -- (conA);
\path [line] (A12) -- (conA);
\path [line] (A13) -- (conA);
\path [line] (B11) -- (conB);
\path [line] (B12) -- (conB);
\path [line] (B13) -- (conB);

\path [line] (conA) -- (Af);
\path [line] (conB) -- (Bf);

\node [main] at (14, 4) (final) {$ \widehat{y} _{ui}$};

\path[line,rounded corners] (Af) -- ++(2.4,0,0) -- (final.south);
\path[line,rounded corners] (Bf) -- ++(2.4,0,0) -- (final.north);
\end{tikzpicture} 
}
\caption{The proposed framework with bi-parties graph and Weighted influence users as additional input.}
\label{fig:proposedmodel}

\end{figure}

\subsection{Prediction and optimization}
After a number of propagation iterations, the embedding vector $E^*$ will be acquired, the predicting score between user $u_i$ on item $i_j$ can be calculated by $\widehat{y}_{ui}={e_{u_i}^*}^\top  e^*_{i_j}$. We build the loss function with Bayesian Personalized Ranking because it is the best suitable method from implicit feedback datasets \cite{rendle2012bpr}. With two pooling observable sets, we have $\Omega^+_{ui}$ is observed relations and $\Omega^-_{uj}$ is unobserved relations. Thus,

\begin{equation}
    Loss_{bpr} =  \sum_{\Omega^+_{ui}}\sum_{\Omega^-_{uj}} -ln \sigma ( \widehat{y}_{ui} - \widehat{y}_{uj} ) +  \lambda  \parallel  \Phi   \parallel ^2_2
\end{equation}

where $\Phi$ is embedding $E^*$ and all of trainable weight $W_1, W_2$, bias $b$; $\sigma(.)$ is sigmoid function and $\lambda$ controls the regularization.

We generalize the propagation process from the two input matrices, through the output embedding and stack layers to calculate a predicted rating value in Figure \ref{fig:proposedmodel}.

\section{Experiments}
We make experiments with our proposed model on three well-known datasets which are Gowalla, Amazon-book and Yelp2018; and compare the result with three state-of-the-art models which are GCMC, NGCF and LightGCN.
\subsection{Datasets Description}
We use setting 10-core with all datasets to ensure that each user has at least ten interactions. For each dataset, 20\% of interactions were random selected as test set and the remaining (80\% interaction) is used for training process. We present a summary of the data sets in \textbf{Table \ref{tabStatic}}.
\begin{itemize}
    \item \textbf{Gowalla}: Gowalla was a location-based social networking service and users were able to check in at their current location.
    \item \textbf{Amazon-book}: This dataset contains product reviews and metadata from Amazon books.
    \item \textbf{Yelp2018}: Yelp is a popular online directory for discovering local businesses such are hotels, bars, restaurants, and cafes.
\end{itemize}
\begin{table}
\centering
\caption{Statistic of the experiment datasets.}\label{tabStatic}
\begin{tabular}{|l|l|l|l|l|}
\hline
Dataset &  \#Users & \#Items & \#Relations & Density\\
\hline
Gowalla & 29,858 & 40,981 & 1,027,370 & 0.000084\\
Amazon-book & 52,643 & 91,599 & 2,984,108 & 0.00062\\
Yelp2018 & 31,668 & 38,048 & 1,561,406 & 0.00130\\
\hline
\end{tabular}
\end{table}
\subsection{Experimental Settings}
\subsubsection{Setting parameters}
We configure our model with embedding
fixed-size to 64 for all models and the embedding parameters and Weighted matrix $W_1, W_2$ are initialized with the He normal weight method. The optimization process was done with Adam algorithm \cite{kingma2017adam}. In the settings, $3-layers$ give the best results. 
\subsubsection{Baseline.} To demonstrate the result, we compare our proposed model with the following state-of-the-art methods:
\begin{itemize}
    \item \textbf{GCMC}\cite{wang2020multi} designed with two modules, decomposer and combiner, this method distinguishes the underlying ranking motives underlying the clearly observed interactions between the user and the item. The first module decomposes rating values to extract the interactions and the second one combines all signals into a unified embeddings for prediction.
    \item \textbf{NGCF}\cite{NGCF} conducts propagation processes on embeddings with several iterations. The stacked embeddings on output contains high-order connectivity in interactions graph. The collaborative signal is encoded into the latent vectors and it make the model more sufficient.
    \item \textbf{LightGCN} \cite{he2020lightgcn}: focus on the neighborhood aggregation component for collaborative filtering. This model uses linearly propagating to learn users and items embeddings for interaction graph. The final embedding is weighted sum of all learned embeddings.
\end{itemize}

\subsection{Results}
The overall performance comparison shows in \textbf{Table \ref{tabResult}}. Our proposed model gives out the better scores, both precision, recall and NGCD.

\begin{table}
\centering
\caption{Overall Performance Comparisons}\label{tabResult}
\begin{tabular}{| c|c|c| c|c|c| c|c|c| c|}
\hline
 \textbf{Dataset} & \multicolumn{3}{|c|}{\textbf{Gowalla}} & \multicolumn{3}{|c|}{\textbf{Amazon-book}} & \multicolumn{3}{|c|}{\textbf{Yelp2018}}\\
   & precision & recall & ndcg@20 &  precision & recall & ndcg@20 & precision &  recall & ndcg@20 \\ 
\hline
GCMC & 0.03690 & 0.11799 & 0.09042 & 0.01122 & 0.02539 & 0.02033 & 0.02320 & 0.05114 & 0.04141 \\
NGCF & 0.04080 & 0.13123 & 0.11149 & 0.01713 & 0.41160 & 0.03045 & 0.02192 & 0.04865 & 0.03917\\
LightGCN & 0.03961 & 0.12637 & 0.11032 & 0.01181 & 0.02657 & 0.02114 & 0.01975 & 0.04262 & 0.03462\\
\hline
\hline
\textbf{Our model}   & 0.04167 & 0.13503 & 0.11648 & 0.01772 & 0.04335 & 0.031619 & 0.02225 & 0.04894 & 0.03954\\
\hline
\end{tabular}
\label{figresult}
\end{table}

The only worse case is when compared to GCMC with the Yelp2018 dataset. We suspect that Yelp2018 clustered users into categories database, which is an advantage for GCMC, as a cluster becomes a component in decomposition and combination. In this case, NGCF and LightGCN also give worse results than GCMC.

\begin{figure}[ht] 
\centering
\scalebox{.7}{
\begin{tikzpicture}
\begin{axis}[
scaled y ticks=real:1000,
ytick scale label code/.code={},
ymax = 120,
symbolic x coords={1, 5,10,15,20,25,30,35,40},
xtick=data,
height=10cm,
width=14cm,
grid=major,
xlabel={Epoch},
ylabel={Loss},
legend style={
cells={anchor=east},
legend pos=north east,
}
]
\addplot coordinates {(1,388.19) (5,95.43) (10, 62.59)  (15, 45.75)  (20, 35.31) (25, 28.65) (30, 23.91) (35, 20.13) (40, 18.89)};

\addplot coordinates {(1, 375.4) (5,93.7) (10, 61.34)  (15, 45.73)  (20, 35.9) (25, 28.31) (30, 24.62) (35, 22.28) (40, 21.02)};

\addplot coordinates {(1, 470.7) (5,120.9) (10, 81.29)  (15, 67.46)  (20, 58.8) (25, 53.21) (30, 49.2) (35, 45.56) (40, 42.17)};

\addplot coordinates {(1, 387.5) (5,105.29) (10, 69.19)  (15, 54.62)  (20, 47.79) (25, 39.21) (30, 35.2) (35, 30.15) (40, 28.17)};
\legend{$WGCN-Gowalla$,$NGCF-Gowalla$, $LightGCN-Gowalla$, $GCMC-Gowalla$}
\end{axis}
\end{tikzpicture}
}
\caption{Test performance of each epoch of MF and NGCF.}
\label{fig:reduceofloss}

\end{figure}
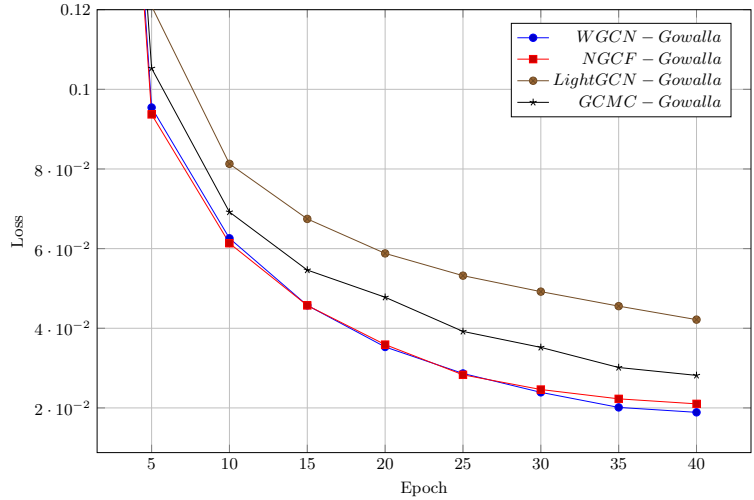

\textbf{Figure \ref{fig:reduceofloss}} shows the comparison of all methods in loss by epoch. Our proposed model reduces the loss faster others. It shows that the propagation process that has been effective recorded after each mini batch.
\section{Conclusion and future works}

\subsection{Conclusion}

In this work, we have proposed a method of weighting the influence among users, which can be considered equivalent to the Social Relation of Trust in the item recommendation problem. The input complement matrix has been calculated to enrich the interactions between the user and the item, and it serves as a moderator of the signals during the extraction and synthesis of messages. By contributing a machine learning model, we believe we have created an inspiration for future studies on the Recommender systems.

\subsection{Future works}
Trust weight graphs between users , which were prepared and send into the proposed computational model, based on the interactions of many users on the same item. This is not a graph that represents real social relationships between users. Future works might be to find out if a social relation graph, by replace with (either collaborative filtering with) the weighted trust graph, will result in better recommendation results. The problem of clustering users according to interests, habits or behaviors also needs to be explored for more effective collaborative filtering.

Last but not least, the tuning parameters for the learning model are difficult to determine optimally, as well as the problem of initializing the embedded layer value at the first step that needs to be considered. We hope that the proposed WGCN model has provided a perspective on data preparation and learning methods from multiple sources of information.
%
%
%
%

\end{document}